# PromptPilot: Improving Human-AI Collaboration Through LLM-Enhanced Prompt Engineering

*Completed Research Paper*

**Niklas Gutheil**[1]
University of Bayreuth
Bayreuth, Germany
niklas.gutheil@uni-bayreuth.de

**Valentin Mayer**[1]
University of Bayreuth
Bayreuth, Germany
valentin.mayer@uni-bayreuth.de

**Leopold Müller**[1]
University of Bayreuth
Bayreuth, Germany
leopold.mueller@uni-bayreuth.de

**Jörg Römmelt**[1]
FIM Research Center for Information Management
Augsburg, Germany
joerg.roemmelt@fim-rc.de

**Niklas Kühl**
University of Bayreuth, Fraunhofer FIT
Bayreuth, Germany
kuehl@uni-bayreuth.de

## Abstract

*Effective prompt engineering is critical to realizing the promised productivity gains of large language models (LLMs) in knowledge-intensive tasks. Yet, many users struggle to craft prompts that yield high-quality outputs, limiting the practical benefits of LLMs. Existing approaches, such as prompt handbooks or automated optimization pipelines, either require substantial effort, expert knowledge, or lack interactive guidance. To address this gap, we design and evaluate PromptPilot, an interactive prompting assistant grounded in four empirically derived design objectives for LLM-enhanced prompt engineering. We conducted a randomized controlled experiment with 80 participants completing three realistic, work-related writing tasks. Participants supported by PromptPilot achieved significantly higher performance (median: 78.3 vs. 61.7; $p = .045$, $d = 0.56$), and reported enhanced efficiency, ease-of-use, and autonomy during interaction. These findings empirically validate the effectiveness of our proposed design objectives, establishing LLM-enhanced prompt engineering as a viable technique for improving human-AI collaboration.*

[1] All authors contributed equally in a shared first authorship.

# Introduction

Large Language Models (LLMs), such as the GPT-series, have emerged as powerful tools and are increasingly being deployed as conversational agents across various domains (Akpan et al., 2025; Bommasani et al., 2021). These advancements have particularly captured public attention, making artificial intelligence (AI) accessible to a broader audience, including non-AI experts. While this democratization expands AI accessibility, it also raises expectations for performance on both individual and societal levels (Noorman & Swierstra, 2023). Conversational agents hold the potential to transform isolated AI applications into interconnected ecosystems of multiple applications, helping to finally fulfill the high expectations placed on AI technologies (Dingler et al., 2021; Wilson, 2022). This transformative potential is particularly evident in LLMs (Deng et al., 2023), functioning as conversational interfaces to influence strategic decision-making (Changeux & Montagnier, 2024).

However, realizing the full potential of LLMs hinges on users' ability to interact with these tools effectively by prompting and appropriately relying on the systems' outputs (Schulhoff et al., 2024; Zamfirescu-Pereira et al., 2023). Despite the increased accessibility of LLMs to non-expert users, crafting prompts that yield high-quality responses often necessitates a certain level of expertise (Zamfirescu-Pereira et al., 2023). Recent work emphasizes the significant impact of prompt engineering on the accuracy and utility of LLM outputs (Amatriain, 2024; Chen et al., 2024; Heston & Khun, 2023). Yet, many users still struggle with effective prompt engineering, limiting their ability to harness the power of LLMs fully (Woo et al., 2024). Therefore, researchers are calling for more empirical validation in real-world settings on how to improve the prompting behavior of users (Chen et al., 2024; H. Liu et al., 2023). Previous prompt improvement techniques, like prompt optimization pipelines, prompt engineering, or prompting handbooks, have limitations in providing users with both systematic and dynamic frameworks to formulate better prompts or requiring users' effort to practice these techniques (Dell'Acqua et al., 2023; Kojima et al., 2022; Zamfirescu-Pereira et al., 2023). In summary, prompt engineering seems to be crucial for working efficiently with LLMs; however, it is complex for users to practice it without proper guidance. Thus, we pose the following question:

*How can we design an LLM-based prompting assistant to improve the effectiveness in human-AI collaboration?*

This paper presents the development of the IT artifact PromptPilot using Design Science Research (DSR) (Peffers et al., 2007) to assist users in crafting more effective prompts to achieve higher quality task performance when collaboratively working on tasks. PromptPilot provides guidance on the current state of a user's prompt, offering suggestions for enhancement based on established prompt engineering principles. If the current prompt does not meet a set of predefined standards, the user is asked more questions to enhance the prompt. We conduct a randomized experiment to evaluate the quality of the output that users achieved with PromptPilot. The code for PromptPilot is publicly available[2].

The results show that the treatment group, which receive support from PromptPilot, consistently outperform the control group regarding task outcomes. This indicates that the PromptPilot meaningfully enhances participants' ability to engineer effective prompts, ultimately leading to higher task performance and therefore improved human-AI collaboration.

By developing and evaluating PromptPilot, we make a theoretical contribution by adding a prompt improvement technique for human-AI collaboration to existing techniques, such as prompt optimization pipelines, prompting handbooks and traditional prompt engineering, by introducing LLM-enhanced prompt engineering. Additionally, this study provides crucial knowledge on the design of LLM-based prompting assistants. The derived and evaluated design objectives (DOs) ensure to improve the performance of the treatment group, which serves as an important base for the further exploration of LLM-enhanced prompt engineering assistants.

The remainder of this paper is structured as follows. We begin by defining key concepts and identifying the research gap in existing literature. Next, we detail the methodical framework of our DSR approach, which was employed to develop the prompt-enhancing LLM-based prompting assistant as a technical artifact.

[2] https://github.com/FraunhoferFITBusinessInformationSystems/PromptPilot

Then, we analyze the artifact's impact on user performance, examining the quality of task outcomes through a user study with 80 participants. Finally, we discuss our results before concluding our study.

## Related Work

The advent of LLMs marks a significant turning point in integrating AI within the workplace, particularly in knowledge-intensive domains (Dell'Acqua et al., 2023). These systems have expanded the capabilities of machines, allowing them to perform tasks traditionally considered the exclusive domain of humans, such as writing, analysis, and creative endeavors (Bommasani et al., 2021; Eloundou et al., 2023). In domains such as marketing, content creation, and customer service, LLMs have already assumed responsibilities previously reserved for human professionals, including text generation, response to inquiries, and drafting communication. This shift can lead to significant productivity gains, as demonstrated by Noy and Zhang (2023), who found that using ChatGPT in professional writing tasks led to a 40 % reduction in time spent on tasks and an 18 % improvement in output quality. Other studies, such as those by Brynjolfsson et al. (2025), propose an increase in productivity of up to 34% overall. However, studies show that some knowledge workers, especially those with high tenure time or high seniority, do not experience increases in productivity through the integration of LLMs (Brynjolfsson et al., 2025; Wang et al., 2023). Therefore, an increase in productivity cannot be taken for granted.

Tackling this issue, Dell'Acqua et al. (2023) provide a prompting handbook to increase overall productivity. This approach instantiates the provision of prompting techniques to users, which existing literature defines as blueprints that describe how to structure a prompt or a dynamic sequencing of multiple prompts (Schulhoff et al., 2024). While many prompting techniques exist, only a small subset are commonly used in research and industry, such as one-shot and chain-of-thought prompting (Schulhoff et al., 2024). The idea of providing prompting techniques to users holds the potential to increase the efficiency and quality of the LLMs' output. However, it takes users time and effort to study and practice these techniques.

More commonly mentioned, the term prompt engineering is typically defined as an iterative process of developing a prompt through modifications or alterations to the prompting technique, which promises users an increase in the LLM's output quality (Schulhoff et al., 2024). Prompt engineering is taught in several organizations to help knowledge workers use LLMs. However, users with less digital skills often do not understand the functionalities of LLMs and prompt engineering, which leads to a lack of effect (Zamfirescu-Pereira et al., 2023). In conclusion, lower performance leads to lower technology adoption rates (Venkatesh et al., 2012).

As a third option, users' performance can be targeted through prompt optimization pipelines, such as abstraction layers that generate interim prompts for the model, thus refining the responses generated by the LLMs (Kojima et al., 2022). Characterized by not relying on additional user input, to an extent, these approaches might lessen the drop in response quality for bad prompts since a broader spectrum of prompts will lead to desired responses. However, the approaches do not necessarily provide a framework to help users formulate better prompts. Additionally, subsequent studies on the effectiveness of prompt optimization pipelines do not sufficiently address whether these tools improve performance compared to using a plain LLM (H. Liu et al., 2023).

We propose that there is a need for an artifact that helps users systematically improve their prompts by providing them with prompt-specific information through an interface. This study stands out from existing literature by aiming to design an LLM-based prompting assistant to enhance the quality of LLM interactions by improving user prompts with LLM-generated suggestions. This artifact serves as an abstraction layer for generating a final prompt while simultaneously offering feedback on how prompts can be optimized. The primary goal of this prompting assistant would be to solve work-related tasks with higher quality than with a plain LLM, enabling us to make suggestions on the viability of such a tool and subsequently setting the stage for further research.

## Method

We employ the DSR process to address our research question (Gregor & Hevner, 2013). DSR is a problem-solving approach that entails developing novel IT artifacts, thereby addressing practical issues while simultaneously creating generalizable design knowledge (Hevner et al., 2004; March & Smith, 1995). This iterative process involves two core activities: building and evaluating an artifact. The build phase centers on designing and creating the artifact, whereas the evaluation phase focuses on assessing its effectiveness against relevant criteria (March & Smith, 1995). In developing our artifact, we merge the build and evaluate activities within the established six-phase DSR framework proposed by Peffers et al. (2007), which consists of problem identification, definition of DOs, design and development, demonstration, evaluation, and communication. Figure 1 illustrates our implementation of these six DSR phases, concisely summarizing the activities undertaken in this study.

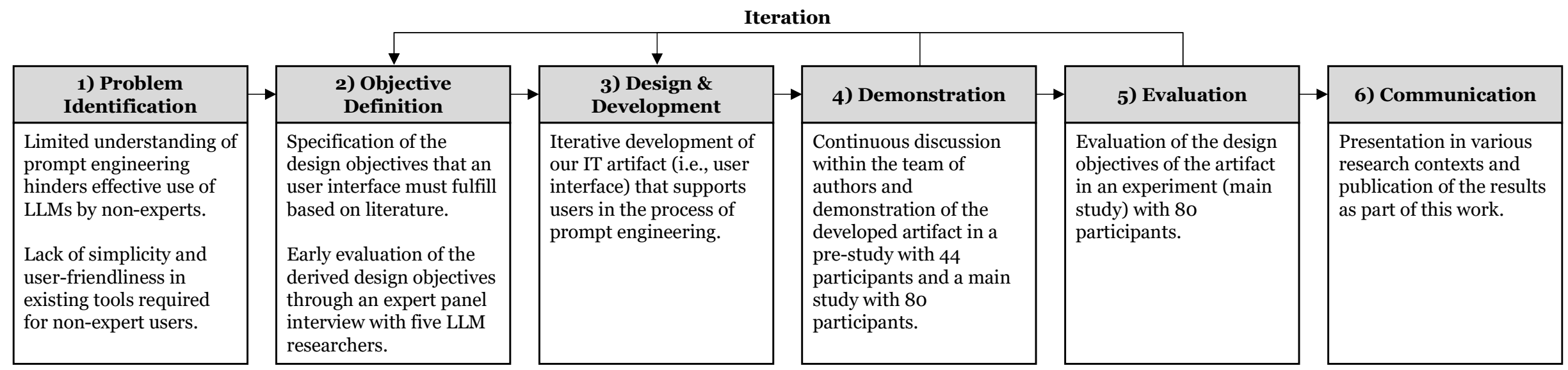

**Figure 1. Overview of Our DSR Process Based on Peffers et al. (2007)**

In **step 1** of our DSR method, we conducted a literature review to motivate and identify our problem. This involved synthesizing existing knowledge in the field of prompt engineering, prompting techniques, and prompt optimization. **Step 2** involved deriving DOs from the descriptive knowledge base established in step 1. Following the DSR approach, DOs were formulated to specify the desired characteristics of a new artifact capable of addressing previously unmet problems (Peffers et al., 2007). Consequently, our DOs served as a foundation for guiding the design and development of the artifact, while also facilitating validation during the demonstration and evaluation phase. We then evaluated and further developed the DOs in a panel interview with five experts in the field of LLMs who are not part of the author team. In **step 3**, the artifact was designed and developed through a systematic design process. In this work, the artifact is an LLM-based prompting assistant that supports the prompt engineering process of users interactively. In **step 4**, the resulting method was demonstrated and documented. Throughout the demonstration phase, an iterative refinement loop was employed to identify and address any shortcomings in the artifact. Therefore, in the first demonstration iteration, we tested the prompting assistant in a pre-study with 44 users to identify issues. This involved revising the DOs and the design and development of the artifact itself, as necessary, to ensure that it effectively addressed the research problem. In the second iteration of step 4, we demonstrated the artifact in our main study with 80 users. In **step 5**, our artifact underwent the evaluation framework of Sonnenberg and vom Brocke (2012), which outlines four essential activities (EVAL 1-4): (1) relevance, (2) design objectives, (3) evaluation criteria, and (4) congruence with real-world applications. As part of these evaluations, we thoroughly assessed our IT artifact, ensuring that it met the necessary standards. Finally, in **step 6** of our DSR approach, we communicate our research findings through the publication of this paper. By making these results publicly available, we aimed to contribute meaningfully to ongoing research initiatives in the field. The communication strategy sought to enhance the visibility of our approach to enhance prompt engineering through an interactive prompting assistant and therefore to increase the output quality and efficiency of work complemented by LLMs. Furthermore, this publication is intended to establish a foundation for future research endeavors, enabling scholars and practitioners to build upon and extend our findings.

To ensure that we met these criteria, we conducted a randomized experiment as part of our main study with a total of 80 participants. We hosted the experiment on *Prolific*, with an equal split between the control and treatment groups. Participants were tasked to solve a set of assignments with the assistance of an LLM, while their behavior was recorded and subsequently registered into a database upon finishing the study. We registered the submitted prompts and their subsequent response to conclude our findings. This is accompanied by suggestions and prompt proposals.

## *Design Objectives*

We have formulated four design objectives (DO1-DO4) to facilitate the design and evaluation of our IT artifact that should support users in the process of prompt engineering. Each of them is backed by literature, next to the motivation outlined in the related work section. In line with Peffers et al. (2007), our DOs are intended to specify how the developed artifact will support a solution to the identified problem, thereby guiding its design and evaluation process.

***DO1: Indicate improvement potential within a specific error domain:*** *The artifact should provide concise and transparent feedback regarding the need for improvement of a prompt within a specific error domain, like missing target audience or purpose of request. Thus, users should be enabled to rapidly understand the error domain without incurring effort expenditure to arrive at a conclusion.*

Considering the vast complexity of LLMs and their nuances in terms of responding, developing an awareness of the quality of a prompt is nontrivial. Especially non-expert users often suffer from a lack of prompting performance, attributable to low experience prompting LLMs (Zamfirescu-Pereira et al., 2023). Kraljic and Lahav (2024) show that it helps users to incorporate their intentions and domains when the system interacts with them in a language-based way. Thus, brief information about the error domain may help the user understand the suggestions for improvement for the current prompt in a given task.

***DO2: Provide goal-oriented guidance for improvement:*** *After identifying the improvement domain, the artifact should provide users with goal-oriented guidance to enhance the prompt. The instructions should be clear and facilitate easy comprehension. To optimize the process, the artifact should leverage automation wherever feasible and practical, e.g., by proactively demanding certain information.*

The lack of prompting knowledge is a key challenge in prompt improvement, as the process remains largely trial and error (Dang et al., 2022). Existing prompt design principles are often too vague to offer actionable guidance (Dang et al., 2022; Schulhoff et al., 2024; Thakur, 2024). Additionally, overly complex explanations overwhelm users, making it difficult to apply improvements effectively (Zamfirescu-Pereira et al., 2023). Furthermore, automation significantly enhances prompt refinement, as AI-assisted tools can identify issues and suggest improvements, reducing reliance on trial and error while maintaining user control (Joshi et al., 2025). Systems like PromptWizard refine prompts dynamically, helping users iterate efficiently while ensuring transparency (Agarwal et al., 2024). Thus, precise and automated guidance enables more effective prompt engineering.

***DO3: Signal improvement and completion of the improvement process:*** *After refining the prompt through targeted and automated feedback, the artifact should signal the point at which the refinement process of a given prompt can be terminated, ensuring that further refinements do not introduce unnecessary complexity or reduce overall quality.*

Prematurely stopping prompt refinement and endlessly optimizing it can both be problematic. Research shows that users often declare a prompt as finished too early, without ensuring its robustness across different cases (Zamfirescu-Pereira et al., 2023). Conversely, the same study observed that participants sometimes hastily added new instructions or completely discarded previous ones when they failed. One guideline is not to specify more conditions than necessary, as unnecessary words or contradictory requirements tend to confuse the model (V. Liu & Chilton, 2022). To address this, the system should signal when a prompt is sufficiently optimized, preventing both under- and over-refinement.

***DO4: Ensure user autonomy in prompt refinement:*** *The assistant must not impose constraints on the user's ability to modify or refine the suggested prompt. While providing structured feedback and recommendations, the system should maintain user autonomy, allowing for manual adjustments and creative flexibility in finalizing the prompt.*

While structured feedback and recommendations can enhance prompt quality, the system should ensure user autonomy by allowing manual adjustments and creative flexibility. Westphal et al. (2023) emphasize that restricting user control can reduce trust and hinder engagement with AI systems. According to Killoran and Park (2022) and Usmani et al. (2023) AI systems should empower people instead of restricting them, share decision-making and provide user control in AI systems. Thus, the assistant should ensure user autonomy in modifying or refining the suggested prompt.

### *EVALs*

Before the conduction of our experiment, we evaluated our artifact with two ex-ante criteria (Sonnenberg & vom Brocke, 2012). EVAL 1 aimed to assess the relevance of our artifact ex-ante. This assessment was grounded in our literature review (see related work and method sections) and in a survey before the experiment, where all participants agreed on the relevance of our artifact. EVAL 2, as the second ex-ante evaluation, included the assessment of our DOs. Again, we derived these DOs firstly out of related literature and evaluated them within our survey before the experiment.

After our experiment, we conducted two ex-post evaluations, EVAL 3 and 4, as proposed by Sonnenberg and vom Brocke (2012). With EVAL 3, we assessed whether the artifact met our derived DOs. Additionally, to ensure that our DSR approach contributes effectively to problem-solving in real-world contexts (March & Smith, 1995), it's essential to develop well-founded and useful artifacts (Gregor & Hevner, 2013). We applied the criteria for methods proposed by Peffers et al. (2007) and March and Smith (1995), which include operationality, efficiency, generality, and ease of use. Furthermore, to assess EVAL 4 we asked for the value-added in our survey ex-post regarding real-world prompt engineering methods. In total, we assessed the performance and practical impact of our IT artifact through this multi-stage evaluation process. The statements in the survey can be revised in Table 1.

| Phase | EVAL | Statements in Survey |
|---|---|---|
| Ex-ante Evaluation | 1 | 1. More knowledge on how to formulate prompts will lead to better results when collaborating with LLMs.<br>2. I would like to get supported by formulating my prompts. |
| | 2 | 1. Advice on where my prompt can be improved (e.g. writing style) would help me improve my prompt. (DO1)<br>2. Concise and easy to understand instructions on how to improve my prompt would help me improve my prompt. (DO2)<br>3. Information about modifications made would verify whether modifications to the prompt have led to improvements. (DO3)<br>4. I would like to keep my autonomy in the decision process regarding which prompt is sent to the LLM. (DO4) |
| Ex-post Evaluation | 3 | 1. The advice on where my prompt can be improved (e.g. writing style) helped me to formulate my prompts. (DO1)<br>2. The guided questions to add information helped me formulate my prompts. (DO2)<br>3. The summary helped me to check where and how my prompt had improved with the changes made by the prompt assistant. (DO3)<br>4. Through the always adjustable prompt, I kept my autonomy regarding which prompt is sent to the LLM. (DO4) |
| | 4 | 1. The prompt assistant saved time for me to fulfill the task. (Efficiency)<br>2. The prompt assistant improved the quality of my solution. (Efficiency)<br>3. The interface is easy to use. (Ease-of-use)<br>4. The prompt assistant is applicable to all kinds of tasks. (Generality)<br>5. The prompt assistant provides consistent support throughout the tasks. (Generality)<br>6. I would always like to use the prompt pilot from now on. (Operationality) |

**Table 1. Evaluation Framework According to Sonnenberg and vom Brocke (2012)**

## Study Design

A controlled experiment was conducted to evaluate the effectiveness of the developed artifact in supporting users during prompt engineering (see Figure 2). The study was structured to capture both objective performance outcomes and subjective user experiences. The experiment began with the ex-ante evaluation, during which participants completed a pre-survey assessing the relevance of our artifact (EVAL 1) and evaluating the design objectives (EVAL 2). Following the survey, participants were randomly assigned to either the control group or the treatment group. Those in the control group completed a sequence of three prompt engineering tasks without PromptPilot, relying solely on their knowledge and intuition. In contrast, participants in the treatment group performed the same tasks using the artifact developed in this research. The artifact was designed to guide users through the process of crafting effective prompts, offering structured support and feedback throughout. All participants completed the same set of tasks to ensure comparability between the groups. The task sequence was also random for every participant to avoid having lower-rated tasks due to fatigue. The tasks simulated realistic scenarios of prompt design, with performance measured through a quantitative evaluation based on predefined metrics, such as task completion time and output quality. These metrics enabled an objective comparison of the effectiveness of the artifact. After completing the tasks, the treatment group was asked to complete an ex-post survey to assess their qualitative perceptions of the artifact. This final stage captured subjective evaluations of the DOs after working with artifact as well as of the evaluation criteria (operationality, efficiency, generality, ease of use) by Peffers et al. (2007) and March and Smith (1995). Taken together, the experimental design enabled a comprehensive assessment of the artifact's impact, both in terms of improving user performance and enhancing the prompt engineering experience.

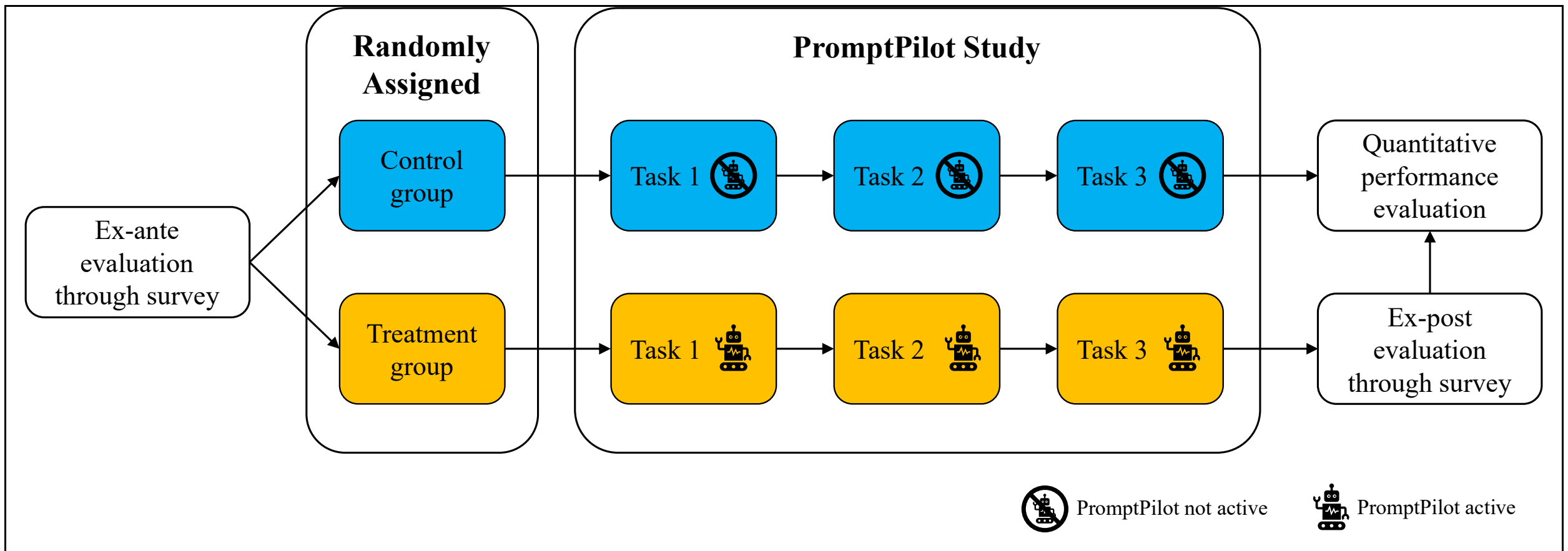

**Figure 2. Study Design of the Experiment**

The study interface is primarily comprised of a panel containing three parts. *Part 1* displays each assignment. Users obtain the information for creating their prompt in this section, but cannot copy the assignment, forcing them to craft their prompts manually. The study interface only accepts the approach of one-shot prompting to ensure that users cannot edit their input and the LLM's response in retrospect. The *second part* represents a chatbot user interface and a prompt input field. The treatment group differs from the control group regarding PromptPilot. In detail, the treatment group is provided with guiding questions that help to optimize the prompt (DO1 & 2, see Figure 3). Afterwards, a short summary informs the user about the improvements made. The information that further editing is final indicates the completion of the improvement process (DO3, see Figure 3). The suggestions offer options to improve the user's prompt based on and comprised of the prompt's structure, specificity, and language. Finally, the suggested prompt uses advanced prompting techniques and improvement indications to help users evaluate the quality of the new prompt. The suggested prompt can still be edited by the user to ensure user autonomy (DO4, see Figure 3). *Part 3* hosts the final answer. To navigate to the next task, the user must fill in the empty box in order to press the "next"-button. Figure 3 shows the user interface in detail with all three outlined parts including the features of PromptPilot. LLaMa 3.1 70B serves as language model for the PromptPilot as well as the language model that finally receives the prompt and solves the task.

To evaluate the responses generated by participants, we employed an LLM-as-a-judge approach, a method previously validated to closely align with human preference ratings (Zheng et al., 2023). Specifically, GPT-4o served as an impartial evaluator, utilizing LangChain's *LabeledScoreStringEvalChain* to systematically assess each response. The evaluation was based on five criteria derived from the LangChain framework: helpfulness, relevance, correctness, depth, and level of detail. Responses were scored on a scale from 1 (very poor) to 100 (excellent). Each participant's response was compared to a high-quality benchmark answer that had been previously generated using GPT-4.5 and manually verified by the research team. Additionally, GPT-4o provided explanatory rationales alongside each numerical score. To ensure the reliability of this automated evaluation, we manually inspected a random subset of the score–rationale pairs, confirming their consistency with both the benchmark solutions and our qualitative expectations.

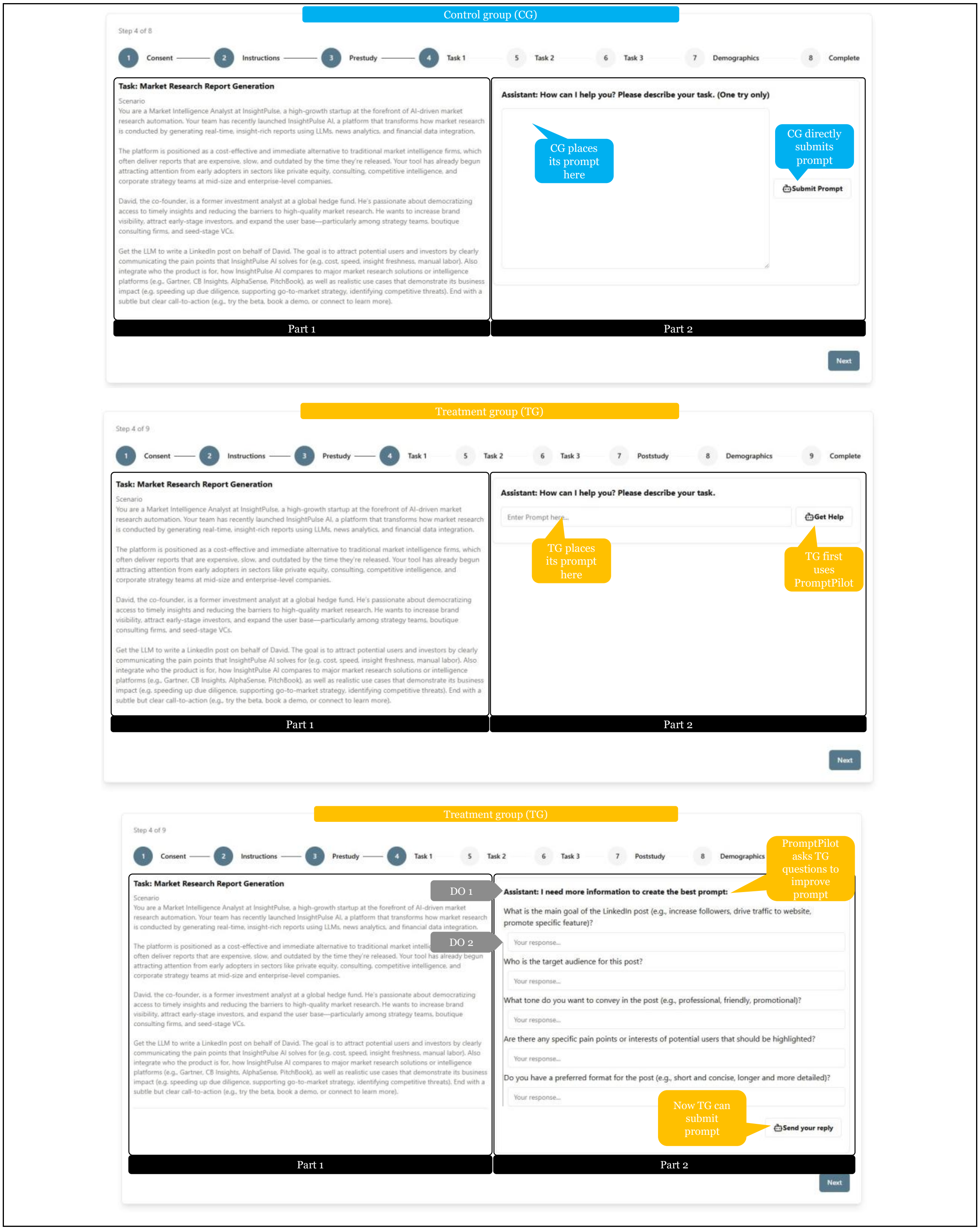

Control group (CG)
Step 4 of 8
1 Consent
2 Instructions
3 Prestudy
4 Task 1
5 Task 2
6 Task 3
7 Demographics
8 Complete
Task: Market Research Report Generation
Scenario
Assistant: How can I help you? Please describe your task. (One try only)
CG places its prompt here
CG directly submits prompt
Submit Prompt
Part 1
Part 2
Next
Treatment group (TG)
Step 4 of 9
1 Consent
2 Instructions
3 Prestudy
4 Task 1
5 Task 2
6 Task 3
7 Poststudy
8 Demographics
9 Complete
Task: Market Research Report Generation
Scenario
Assistant: How can I help you? Please describe your task.
Enter Prompt here...
Get Help
TG places its prompt here
TG first uses PromptPilot
Part 1
Part 2
Next
Treatment group (TG)
Step 4 of 9
Task: Market Research Report Generation
Scenario
PromptPilot asks TG questions to improve prompt
DO 1
DO 2
Assistant: I need more information to create the best prompt:
What is the main goal of the LinkedIn post (e.g., increase followers, drive traffic to website, promote specific feature)?
Your response...
Who is the target audience for this post?
What tone do you want to convey in the post (e.g., professional, friendly, promotional)?
Are there any specific pain points or interests of potential users that should be highlighted?
Do you have a preferred format for the post (e.g., short and concise, longer and more detailed)?
Now TG can submit prompt
Send your reply
Part 1
Part 2
Next

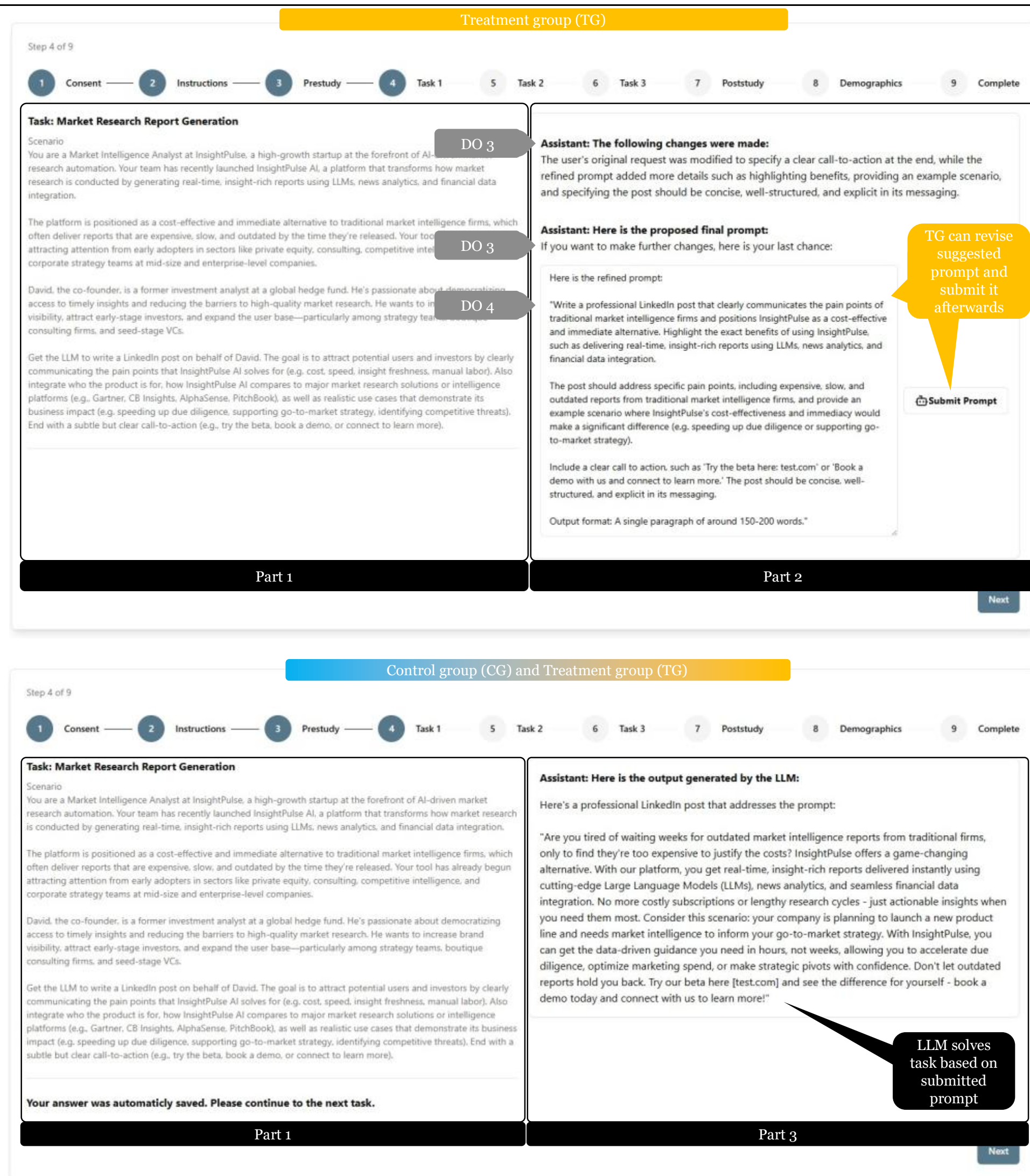

**Figure 3. Study Interface of Control and Treatment Group**

## *Study Tasks*

Since PromptPilot is designed to serve work-related tasks mainly, the assignments should reflect this characteristic. Work-related tasks can entail a plethora of scenarios. This study, therefore, proposes a custom set of assignments specifically tailored to represent work-related activities. Dell'Acqua et al. (2023) show that tasks outside the jagged technological frontier of AI capabilities lead to a decrease in quality when solved by an LLM-assisted knowledge worker. Therefore, the tasks should be well within the capabilities of LLMs to ensure that workers generate high-quality output. We derive our tasks based on the following criteria:

(1) *Any task or assignment is to be work-related:* All tasks or assignments should be relevant to work-related scenarios, reflecting situations that could realistically occur in or with a connection to a professional environment.

(2) *Any scenario in which a task or assignment is embedded, should be unfamiliar to the study participant:* To ensure that the study participant relies on the LLM’s response, it is crucial to prevent situations where the participant has sufficient expertise to assess the quality of the response independently.

(3) *A response to a generic prompt should not align with the ideal outcome if all necessary information has not been provided:* To ensure a distinction between the treatment and control groups, responses from well-crafted prompts should differ from those generated by poorly designed prompts.

(4) *There should be a spectrum of acceptable solutions rather than a single, correct response:* Considering these criteria, this study creates three distinct case scenarios complemented with subsequent assignments. Inspired by consulting case studies, the resulting tasks are modified to adhere to the abovementioned criteria.

Thus, we constructed three tasks of similar complexity and outlined characteristics. In task 1, the user should write a social media thread to attract potential users and investors to a new AI-based market research tool. Task 2 requires the user to create a customer persona for a potential buyer of an eco-friendly care product. Finally, in task 3, the user must write a blog post styled as a short story as a fiction author. In summary, the assignment across all tasks is to generate textual content based on the given information in the task description.

## Results

### *Demographic Distribution*

A total of 80 participants are evenly distributed across the control and treatment groups (n = 40 each). The sample is gender-balanced, with 46% identifying as female and 53% as male. Participants span a wide age range (20-60+), with the largest age groups being 25 to 34 years (42%) and 40 to 49 years (27%) (see Figure 4).

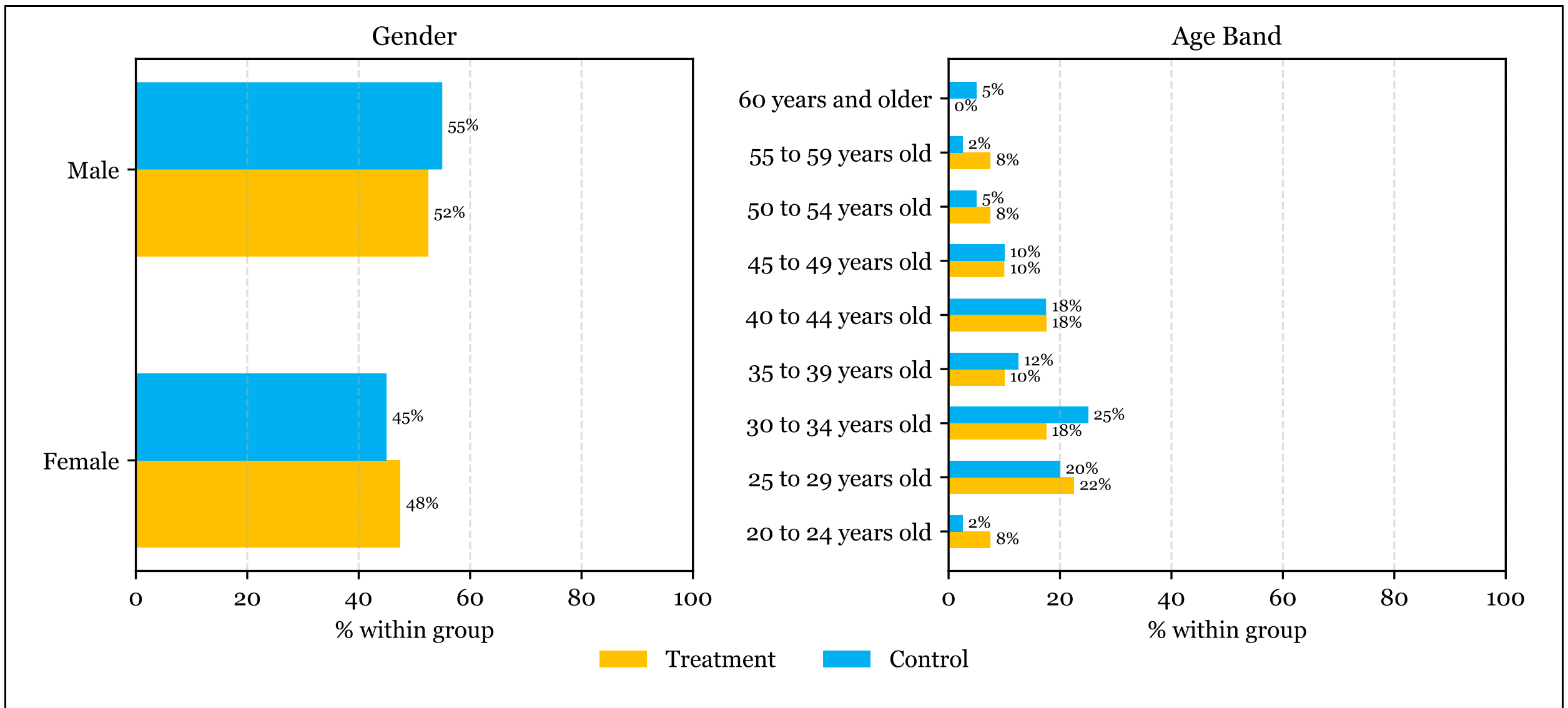

**Figure 4. Demographic Distribution of Experiment Participants**

Regarding prior experience with LLMs, 62% report using LLMs at least once a week, including 12% who used them daily. Professional backgrounds are diverse, with notable representation from IT and software development (20%), healthcare (16%), marketing and sales (16%), and academia or research (11%) (see Figure 5). Figure 5 shows a light imbalance in LLM use frequency between treatment and control group. However, the distribution of LLM use frequency was comparable across treatment and control groups ($\chi^2(4) = 3.18$, $p = .53$), with the largest absolute difference in any usage category being ±12.5 percentage points. This indicates that both groups entered the experiment with a similar baseline experience in LLM usage.

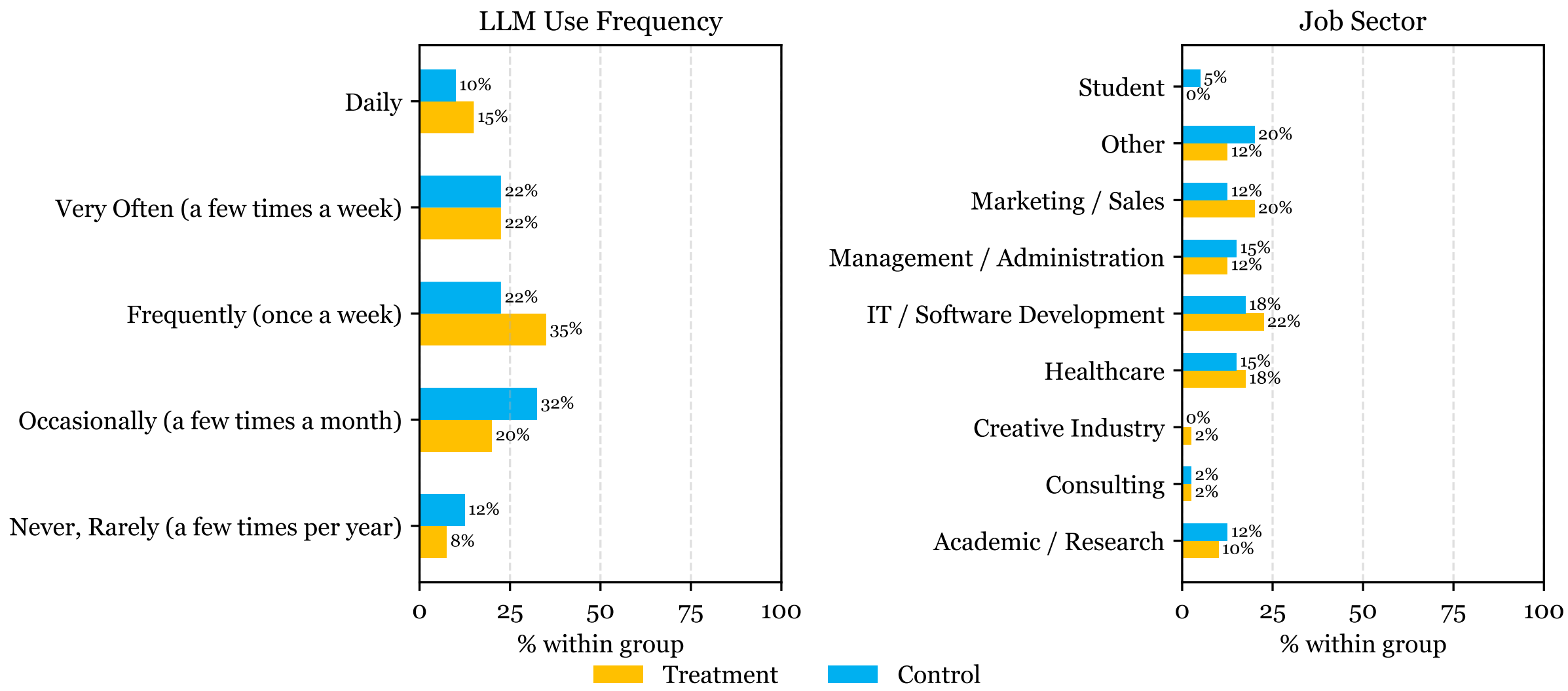

**Figure 5. LLM Experience and Job Sector of Experiment Participants**

## *Performance by Group*

We compared participants' scores across three tasks and the overall mean. The results are visualized in Figure 6, showing consistently higher mean performance in the treatment group across all tasks. The bars represent the mean scores for the treatment and control groups, separately for each task and the aggregated overall mean. The error bars denote one standard deviation.

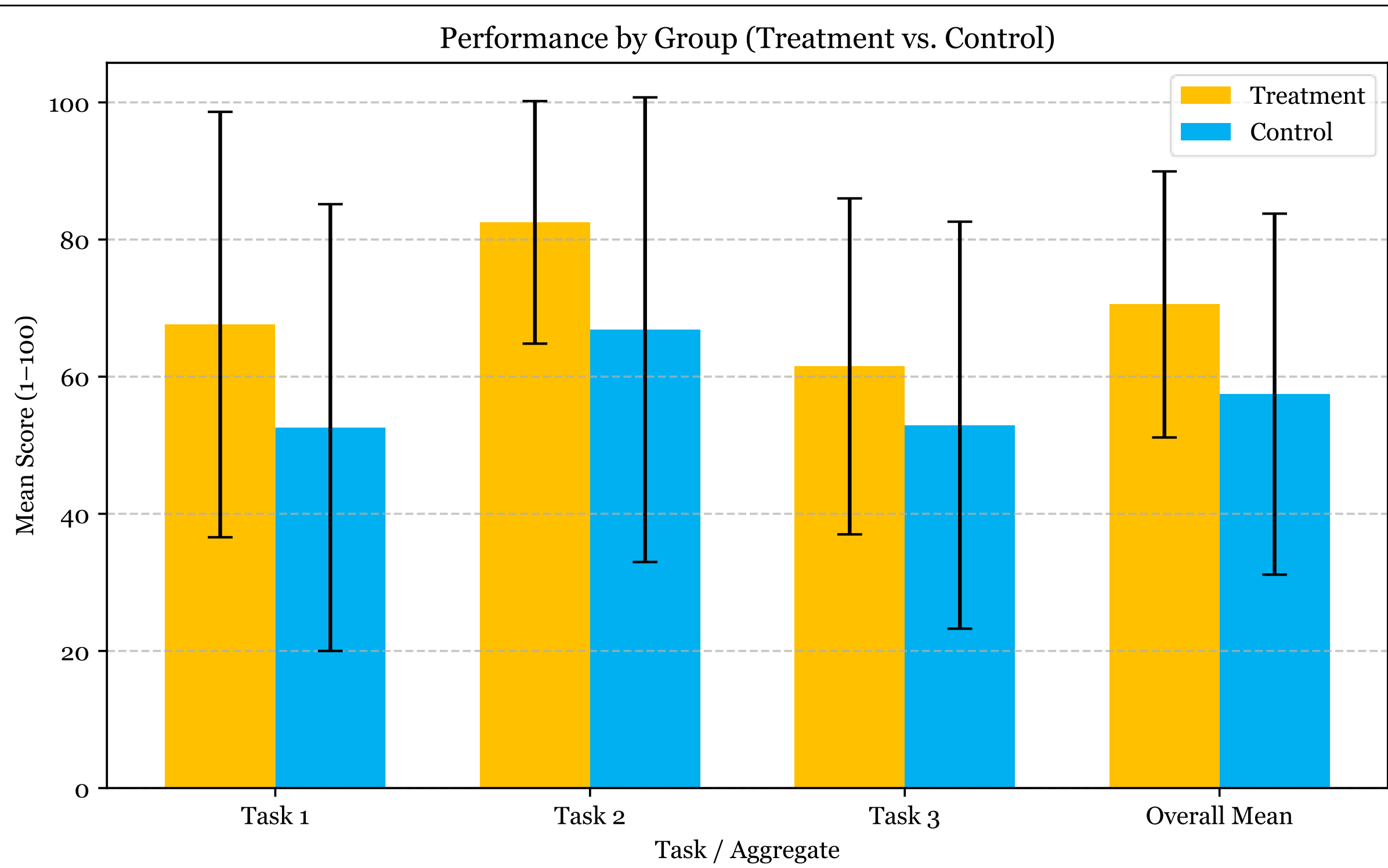

**Figure 6. Performance Comparison Across Different Tasks**

On average, participants in the treatment group score higher than those in the control group across all three tasks and in the overall mean. The mean overall score for the treatment group is 70.53 (SD = 19.40), compared to 57.45 (SD = 26.32) in the control group. Task-specific means for the treatment group are 67.60 (SD = 31.02) for Task 1, 82.50 (SD = 17.69) for Task 2, and 61.50 (SD = 24.50) for Task 3. In the control group, the respective means are 52.58 (SD = 32.58), 66.85 (SD = 33.88), and 52.92 (SD = 29.68).

Participants who used PromptPilot (n = 40) achieve higher overall quality scores than those in the control group (n = 40). The treatment group's median overall score is 78.3 (IQR = 28.1) versus 61.7 (IQR = 44.2) for control. As assumptions of normality were violated (Shapiro–Wilk $p < .05$), we use one-sided Mann–Whitney U tests to compare ranks between groups. The test confirms that this difference is significant, $U$ = 1038, $p$ = .011; after Holm correction, the results remain reliable, padj = .045. The effect is small-to-medium (Cohen's d = 0.56, 95 % CI [0.14, 1.03]).

Task-wise analyses show a significant rank advantage for Task 1 (padj = .045). In contrast, Task 2 (U = 939, one-sided p = .086, padj = .171) and Task 3 (U = 921, p = .123, padj = .171) did not differ significantly after adjustment. Table 2 shows the results of the analyses, with higher scores indicating better quality (scale = 1–100). Medians are reported with inter-quartile ranges in brackets. Holm-adjusted p-values control the family-wise error rate across the four one-sided Mann-Whitney U tests (treatment > control). Further, Cohen's d is Hedges-corrected.

| **Task / Aggregate** | **Treatment Median [IQR]** | **Control Median [IQR]** | **U** | **p (raw)** | **pHolm** | **Cohen's d (95% CI)** |
|---|---|---|---|---|---|---|
| Task 1 | 85.0 [50.8] | 55.0 [65.0] | 1036 | .011 | **.045** | 0.47 [0.03, 0.97] |
| Task 2 | 90.0 [ 7.3] | 90.0 [70.0] | 939 | .086 | .171 | 0.57 [0.16, 1.00] |
| Task 3 | 65.0 [40.0] | 55.0 [60.0] | 921 | .123 | .171 | 0.31 [-0.12, 0.77] |
| **Aggregate** | 78.3 [28.1] | 61.7 [44.2] | 1038 | .011 | **.045** | 0.56 [0.14, 1.03] |

**Table 2. Results of the Task-Wise Analyses.**

In summary, the quantitative analysis of the experiment demonstrates that PromptPilot enhanced the overall performance of the treatment group for all tasks. In addition to the quantitative results, the treatment group completed a qualitative survey regarding the collaboration with PromptPilot, which is outlined in the evaluation section.

## Evaluation

The evaluation process was conducted against four predefined criteria (EVAL 1-4). These criteria encompass assessments of PromptPilot's relevance (EVAL 1), its alignment with the design objectives (EVAL 2), and its performance across multiple evaluation criteria, including efficiency, ease-of-use, generality, and operationality as recommended by Peffers et al. (2007) and March and Smith (1995). Furthermore, we evaluate the value-added contribution of our approach in relation to existing real-world prompt engineering methods (EVAL 3 and 4). A summary of the evaluation results is presented in Figure 7 and 8.

To assess the relevance of our IT artifact, we conducted an initial ex-ante evaluation **(EVAL 1)** based on participants' agreement with two key statements. The results show strong support for the artifact's relevance: the statement "More knowledge on how to formulate prompts will lead to better results when collaborating with LLMs" achieves a high mean score of 4.31 (SD = 0.65), indicating both strong agreement and a consistent response pattern. Similarly, the statement "I would like to get supported by formulating my prompts" receives a mean of 3.86 (SD = 1.00), further underscoring the perceived value of assistance in prompt formulation. Together, these findings confirm the importance of providing more information and support on formulating prompts (see Figure 7).

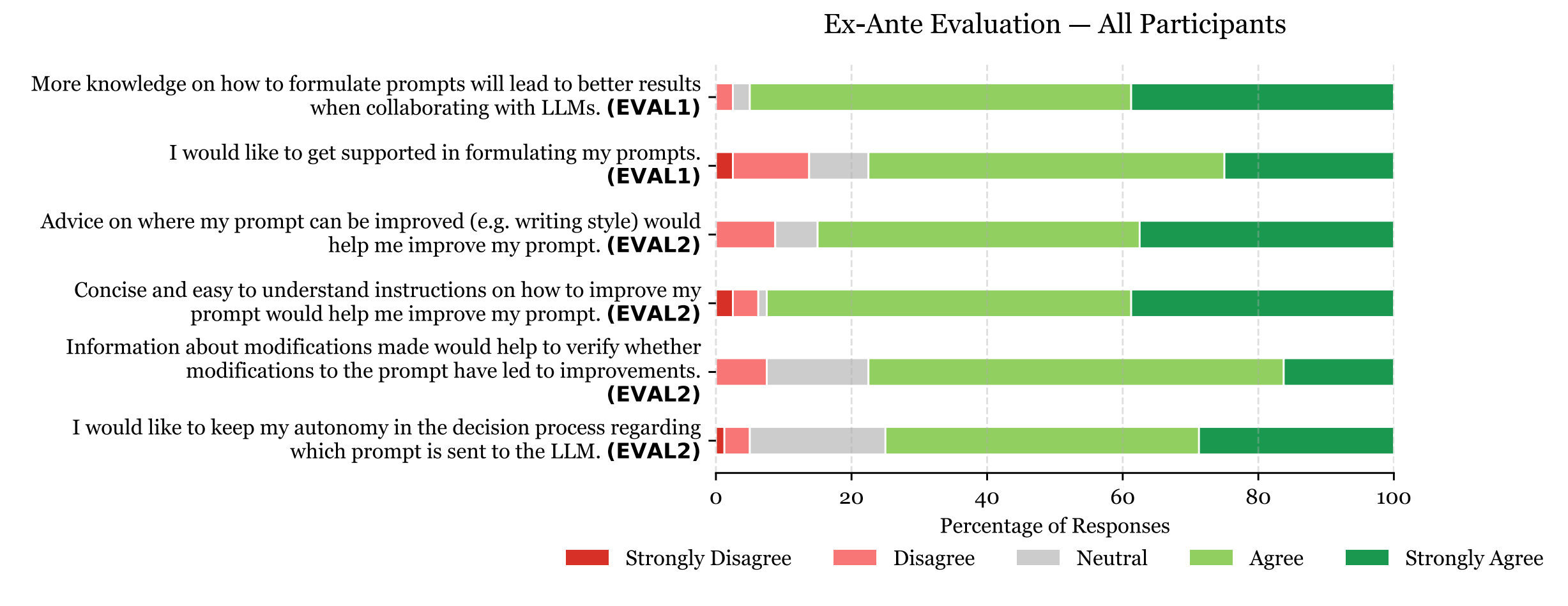

**Figure 7. Results of EVAL 1 and 2**

The second ex-ante evaluation **(EVAL 2)** assesses the alignment of our derived DOs with participants' needs and expectations. The results demonstrate strong support across all objectives. DO1 ("Advice on where my prompt can be improved") receives a mean score of 4.14 (SD = 0.88), and DO2 ("Concise and easy to understand instructions on how to improve my prompt would help me improve my prompt") achieves an even higher mean of 4.22 (SD = 0.86), both indicating high agreement. DO3 ("Information about modifications made would help to verify whether modifications to the prompt have led to improvements") is also positively evaluated with a mean of 3.86 (SD = 0.78). Finally, DO4 ("I would like to keep my autonomy in the decision process regarding which prompt is sent to the LLM") scores a mean of 3.98 (SD = 0.87). These results collectively suggest that the design objectives are well-founded and resonate with the participants' preferences (see Figure 7).

The first ex-post evaluation **(EVAL 3)**, conducted after participants interacted with PromptPilot, confirms the successful implementation of our DOs within the IT artifact. The treatment group reports that evaluating their prompts was helpful for prompt engineering (DO1), with a mean score of 4.28 (SD = 0.55). Support suggestions are particularly well received (DO2), reflected in a high mean of 4.42 (SD = 0.59). Additionally, the summary of improvements made are found useful for verifying improvements (DO3), with a mean of 4.25 (SD = 0.74). Importantly, participants appreciate retaining autonomy in the prompt engineering process (DO4), as indicated by a mean score of 4.33 (SD = 0.62). These results collectively underscore the artifact's effectiveness in supporting users during prompt formulation and improving prompt quality in line with the defined DOs (see Figure 8).

The second ex-post evaluation **(EVAL 4)** assesses the practical effectiveness of PromptPilot in real-world prompt engineering scenarios. Participants report that the tool improved task efficiency, saving time (M = 4.20, SD = 0.88) and enhancing solution quality (M = 4.53, SD = 0.51). The prompt assistant is rated as easy to use (M = 4.15, SD = 0.86), supporting a positive evaluation of its usability. Regarding generality, respondents agree that PromptPilot provided consistent support across different tasks (M = 4.22, SD = 0.70) and rate it as broadly applicable (M = 4.12, SD = 0.88). Finally, the participants declare with a mean score of 4.28 (SD=0.68) that they would always like to use PromptPilot from now on, indicating that its operationality is affirmed and that participants would be willing to continue using PromptPilot in future tasks. These results highlight PromptPilot's efficiency, easy usability, general applicability, and operationality for sustained use in diverse prompting contexts (see Figure 8).

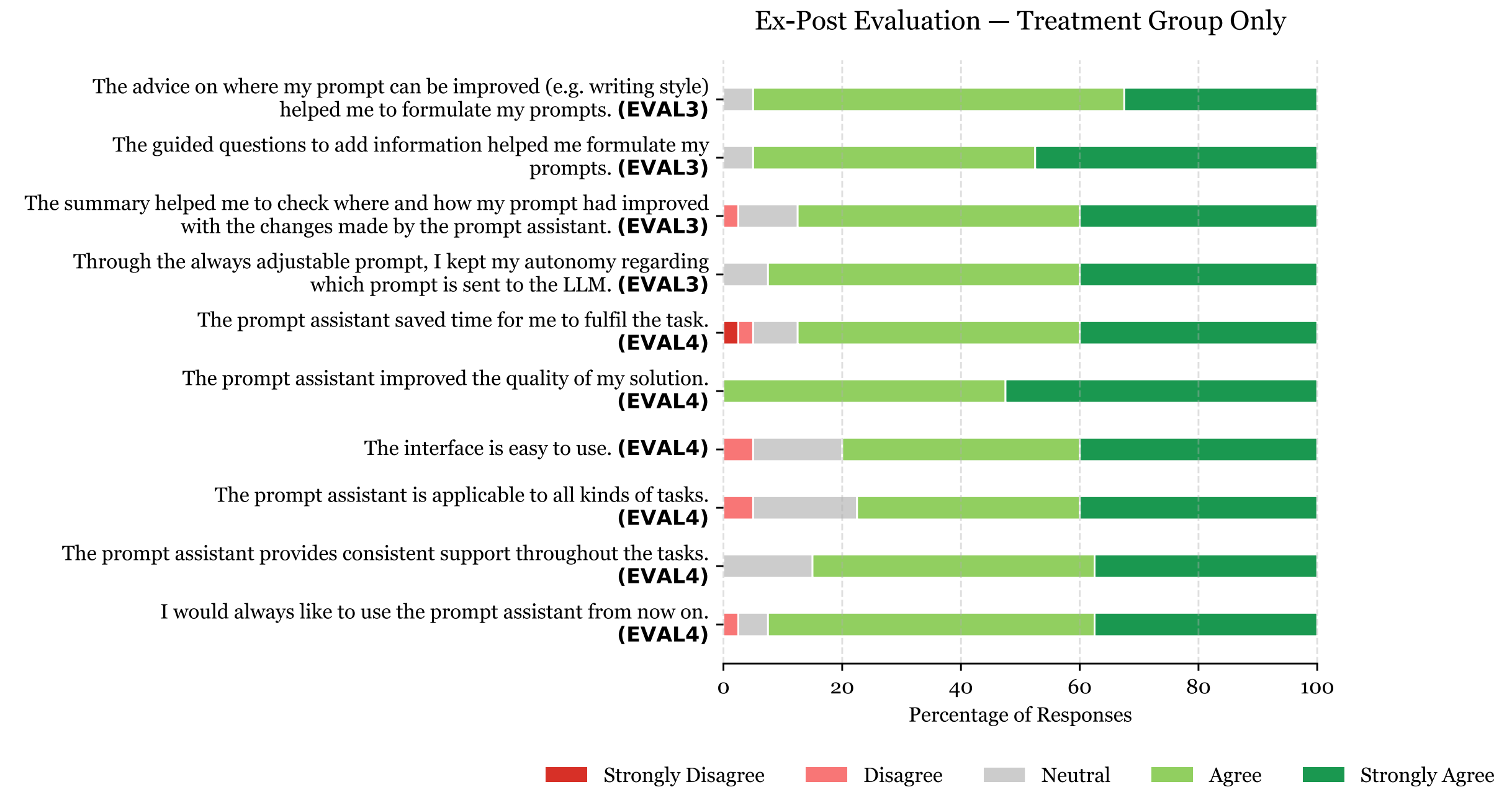

**Figure 8. Results of EVAL 3 and 4**

## Discussion

As shown in the related work section, existing techniques to achieve results in human-AI-collaboration have strong limitations and often lack the desired quantifiable effect. For example, prompt handbooks are complex and take time to study, prompt engineering is often difficult to understand for non-experts, and prompt optimization pipelines often do not actually help users to get better at prompting. Through our research, we aim to design an LLM-based prompting assistant that helps users formulate better prompts and increases the quality of the LLM's output when working on tasks collaboratively. The results of our study show that our IT artifact PromptPilot addresses our research goal by enabling users to produce significantly higher task performance overall by enhancing prompt engineering. Analogous to our experimental results, the users' self-reported statements underscore the positive effect of PromptPilot. In EVAL 4, the perceived quality increase of the solutions stands out, with all users solely agreeing that PromptPilot improved the quality of their solution. With PromptPilot, we synthesize the three existing techniques to counter their weaknesses. Building on the limitations of prompt handbooks, traditional prompt engineering, and prompt optimization pipelines, we add to existing theory by introducing LLM-enhanced prompt engineering as an innovative technique of achieving higher performance when collaborating with LLMs on tasks. LLM-enhanced prompt engineering is easy to use, applicable to a wide range of tasks and users, and leads to higher quality prompts and finally to better task outcomes, as shown in this study. We therefore propose that researchers should build on our results to create additional design knowledge and further enhance the technique of LLM-enhanced prompt engineering with the final goal of improving human-AI collaboration.

Besides the introduction of LLM-enhanced prompt engineering, our study holds relevant design knowledge for researchers. The exploration of improvement domains and clear instruction for the improvement of prompts have been valuable objectives for the design of LLM-based prompting assistants (DO1, DO2) in this study. The ex-ante evaluation indicated that users perceived DO3 and DO4 as medium relevant. However, after working with PromptPilot, the treatment group ranked the same DOs as quite important, with a higher mean than before. This indicates that users saw an improvement notification of the prompt as well as their own autonomy in the optimization process as less relevant before the experiment. The use of PromptPilot convinced users that these two factors are highly relevant, clearly distinguishing our artifact from traditional prompt optimization pipelines, where users get no feedback. Future research should build upon this knowledge and further enhance LLM-enhanced prompt engineering by additional DOs.

Our findings also have implications regarding the European Union's AI Act, in particular Article 4, which requires companies to ensure a sufficient level of AI literacy for their employees when working with AI systems (Regulation 2024/1689). Due to Article 4, competencies such as prompt engineering must be ensured legally by companies to guarantee efficient and ethical use of AI by their employees. At this point, the questions arise whether AI systems like PromptPilot can teach AI literacies like prompt engineering by themselves by interacting with users, as PromptPilot does by supporting prompt engineering, or if the need for AI literacies like prompt engineering is even obsolete for users due to AI assistants making sure to provide humans with the needed literacy. LLM-based assistants could therefore guide users and ensure requested AI literacies by interacting with them during tasks. This could prevent users from making certain mistakes, leading not only to an overall increase in performance, as in this study, but also to more ethical and compliant behavior. This is an important topic that warrants further discussion and research.

For practitioners, our study implies that LLM-enhanced prompt engineering improves the performance of human-AI collaboration. Further, integrating mechanisms such as the DOs of PromptPilot into user interfaces could lead to users sending less prompts to LLMs which can save computing power. Developers and deployers of LLM-based assistants should therefore incorporate our DOs into their work to increase user acceptance and satisfaction but also behave more sustainable when interacting with AI systems. For example, advanced LLM-based assistants like OpenAI's Deep Research already started to ask further questions to the user before executing the required task.

## Limitations and Suggestions for Future Research

Despite our rigorous study design, several limitations offer opportunities for future research. PromptPilot integrates multiple prompting techniques but the individual contributions of these components remain unclear. Future research could systematically evaluate each element to understand its specific impact better and enhance the overall design. Additionally, the effectiveness of PromptPilot varied by task, with tasks 2 and 3 showing weaker, non-significant improvements. Further investigation is needed to identify task characteristics that determine PromptPilot's efficacy, clarifying under what conditions it is most beneficial. In this context, future research should emphasize the mechanisms of tasks and PromptPilot that drive the quality increases. Our comparison exclusively involved scenarios with and without PromptPilot, lacking benchmarks against alternative prompting support tools such as optimization pipelines or prompting handbooks. Future studies should include comparative analyses to position PromptPilot's effectiveness within the broader landscape and highlight possible improvements. Furthermore, our study evaluated only the quality of the output generated, without directly measuring prompt quality itself. Future research should explicitly assess whether and how PromptPilot enhances prompt quality, the progression from initial to final prompts, and the relationship between improved prompts and output quality. Our analysis also omitted productivity metrics such as time-on-task and user effort. Investigating these factors would provide a more comprehensive understanding of PromptPilot's impact on task efficiency and user productivity. Finally, minor group imbalances regarding prior LLM usage could have influenced performance outcomes. Future research could explore how user characteristics, including prior experience and skill levels, moderate PromptPilot's effectiveness. Finally, examining other prompting techniques, such as chained prompting, and conducting longitudinal studies to identify potential learning effects over time, would significantly enrich our understanding of PromptPilot's broader utility and impact.

## Conclusion

This study addresses a crucial challenge in harnessing the power of LLMs through effective prompt engineering. By developing and evaluating PromptPilot that provides users with LLM-generated suggestions to enhance their prompts, we demonstrate a significant improvement in task outcomes. Our results show that the treatment group, supported by PromptPilot, consistently outperformed the control group, highlighting the effectiveness of our approach in empowering users to craft high-quality prompts. We therefore contribute to traditional techniques by introducing LLM-enhanced prompt engineering as a new technique to improve human-AI-collaboration. This contribution has important implications for both theory and practice, as it sheds light on the mechanisms that facilitate successful human-AI collaboration through improved prompt engineering. By providing design knowledge of a well-functioning prompting assistant that supports users in performing effective prompt engineering, we accelerate the adoption and effective use of LLMs, ultimately unlocking their transformative potential to influence strategic decision-making and expand AI accessibility to a broader audience. Our study responds to the call for empirical validation in real-world settings, offering valuable insights for researchers, developers, and practitioners seeking to design more intuitive and effective interfaces that connect AI-driven tools with humans.